\documentclass[prl,aps,twocolumn,showpacs,preprintnumbers,amsmath,amssymb]{revtex4}

\usepackage{graphicx}
\usepackage{dcolumn}
\usepackage{bm}
\usepackage{color}
\usepackage[utf8]{inputenc}

\begin{document}
\def \tr{{\mbox{tr~}}}
\def \ra{{\rightarrow}}
\def \ua{{\uparrow}}
\def \da{{\downarrow}}
\def \be{\begin{equation}}
\def \ee{\end{equation}}
\def \ba{\begin{array}}
\def \ea{\end{array}}
\def \bea{\begin{eqnarray}}
\def \eea{\end{eqnarray}}
\def \nn{\nonumber}
\def \l{\left}
\def \r{\right}
\def \half{{1\over 2}}
\def \etal{{\it {et al}}}
\def \cH{{\cal{H}}}
\def \cM{{\cal{M}}}
\def \cN{{\cal{N}}}
\def \cQ{{\cal Q}}
\def \cI{{\cal I}}
\def \cV{{\cal V}}
\def \cG{{\cal G}}
\def \cF{{\cal F}}
\def \cZ{{\cal Z}}
\def \bS{{\bf S}}
\def \bI{{\bf I}}
\def \bL{{\bf L}}
\def \bG{{\bf G}}
\def \bQ{{\bf Q}}
\def \bK{{\bf K}}
\def \bR{{\bf R}}
\def \br{{\bf r}}
\def \bu{{\bf u}}
\def \bq{{\bf q}}
\def \bk{{\bf k}}
\def \bz{{\bf z}}
\def \bx{{\bf x}}
\def \bpsi{{\bar{\psi}}}
\def \tJ{{\tilde{J}}}
\def \W{{\Omega}}
\def \e{{\epsilon}}
\def \lam{{\lambda}}
\def \L{{\Lambda}}
\def \a{{\alpha}}
\def \t{{\theta}}
\def \b{{\beta}}
\def \g{{\gamma}}
\def \D{{\Delta}}
\def \d{{\delta}}
\def \w{{\omega}}
\def \s{{\sigma}}
\def \f{{\varphi}}
\def \x{{\chi}}
\def \e{{\epsilon}}
\def \h{{\eta}}
\def \G{{\Gamma}}
\def \z{{\zeta}}
\def \hatt{{\hat{\t}}}
\def \hn{{\bar{n}}}
\def \vk{{\bf{k}}}
\def \vq{{\bf{q}}}
\def \gk{{\g_{\vk}}}
\def \nd{{^{\vphantom{\dagger}}}}
\def \yd{^\dagger}
\def \av#1{{\langle#1\rangle}}
\def \ket#1{{\,|\,#1\,\rangle\,}}
\def \bra#1{{\,\langle\,#1\,|\,}}
\def \braket#1#2{{\,\langle\,#1\,|\,#2\,\rangle\,}}


\title{Counterflow superfluid of polaron pairs in Bose-Fermi mixtures in optical lattices}
\author{Ippei Danshita$^{1}$}
\altaffiliation{Present address: Yukawa Institute for Theoretical Physics, Kyoto University, Kyoto 606-8502, Japan}
\author{L. Mathey$^{2}$}
\affiliation{$^1$Computational Condensed Matter Physics Laboratory, RIKEN, Wako, Saitama 351-0198, Japan\\$^2$Zentrum f\"ur Optische Quantentechnologien and Institut f\"ur Laserphysik, Universit\"at Hamburg, 22761 Hamburg, Germany}

\date{\today}

\begin{abstract}
We study the quantum phases of one-dimensional Bose-Fermi mixtures in optical lattices. Assuming repulsive interparticle interactions, equal mass, and unit total filling, we calculate the ground-state phase diagram by means of both Tomonaga-Luttinger liquid theory and time-evolving block decimation method. We demonstrate the existence of a counterflow superfluid (CFSF) phase of polaron pairs, which are composite particles consisting of two fermions and two bosonic holes, in a broad range of the parameter space. We find that this phase naturally emerges in $^{174}$Yb-$^{173}$Yb mixtures, realized in recent experiments,  at low temperatures.
\end{abstract}

\maketitle

The unprecedented control that has been achieved in ultra-cold atom systems in optical lattices has generated a new frontier in exploring quantum phases in the strongly correlated regime. Novel quantum phases have been created in mixtures of different hyperfine states~\cite{weld-09, gadway-10, jordens-08, schneider-08}, atomic species~\cite{catani-08, gunter-06, ospelkaus-06}, and isotopes~\cite{taie-10, sugawa-11} in optical lattices. In these experiments, a wide range of features, such as the statistics of particles, the mass and density ratios, and interparticle interactions, can be precisely varied. This, in turn, has led to the prediction of numerous further exotic phases that can be studied in these systems, including supersolids~\cite{buchler-03, mathey-07-1, hebert-08, mathey-09}, paired superfluids~\cite{mathey-09, kuklov-04, arguelles-08, hu-09}, counterflow superfluids (CFSF)~\cite{ kuklov-04, kuklov-03, lewenstein-04, hu-09, rousseau-08, rousseau-09, eckardt-10}, and Tomonaga-Luttinger liquids (TLL) of polarons~\cite{mathey-04, mathey-07-2}.

Recently, experiments on a Bose-Fermi (BF) mixture of Ytterbium isotopes  ($^{174}{\rm Yb}$-$^{173}{\rm Yb}$)  in optical lattices and the realization of a novel dual Mott insulator have been reported~\cite{sugawa-11}. In this phase the total density $\langle n_{b, i} + n_{f, i}\rangle$, where $n_{b (f), i}$ is the bosonic (fermionic) density at lattice site $i$, is pinned to unit filling, while the individual densities $\langle n_{b, i} \rangle$ and $\langle n_{f, i}\rangle$ are unrestricted and assumed to be close to half-filling. This suggests that the degree of freedom that has to be considered is $c_{f, i}^\dagger c_{b, i}$, where $c_{f,i}^\dagger$ ($c_{b,i}$) is the fermionic creation (bosonic annihilation) operator at site $i$, because the creation of one particle type is matched with the annihilation of the other, thus keeping the density at unity. Thus, the system can be regarded as a liquid of composite fermions~\cite{kuklov-03, lewenstein-04}, which can also be considered as a limiting case of polarons~\cite{mathey-07-1}. If this mixture was either a Bose-Bose (BB) or Fermi-Fermi (FF) mixture, these composite particles are bosonic. For a BB mixture, say, these bosonic particle-hole pairs can condense into a CFSF phase~\cite{kuklov-04, kuklov-03, lewenstein-04, hu-09}. However, the nature of BF mixtures is fundamentally distinct, because the particle-hole pairs are fermionic, and thus cannot function as an order parameter. Rather, these composite fermions have to form pairs themselves to condense.  This constitutes a Bose-Einstein condensation of quartets consisting of two fermions and two bosonic holes, thus creating a polaron-paired CFSF (PP-CFSF) phase.
Previous studies of these systems in the strong-coupling limit have predicted phase separation (PS), Fermi liquid of polarons, spin-density wave (SDW), and CFSF with $p$-wave pairing of the polarons~\cite{kuklov-03, lewenstein-04}. 
This $p$-wave CFSF is fascinating also in the sense that it can be regarded as a topological superfluid~\cite{massignan-10}, which has been extensively discussed in the contexts of liquid $^3$He~\cite{murakawa-11} and the compound of ${\rm Sr}_{2}{\rm Ru}{\rm O}_4$~\cite{kashiwaya-11}. However, in direct simulations of the BF-Hubbard model~\cite{hebert-08, roth-04, takeuchi-05, pollet-06, zujev-08, rizzi-08, mering-08}, the presence of this exotic state has not been confirmed. 

\begin{figure*}[tb]
\includegraphics[scale=0.45]{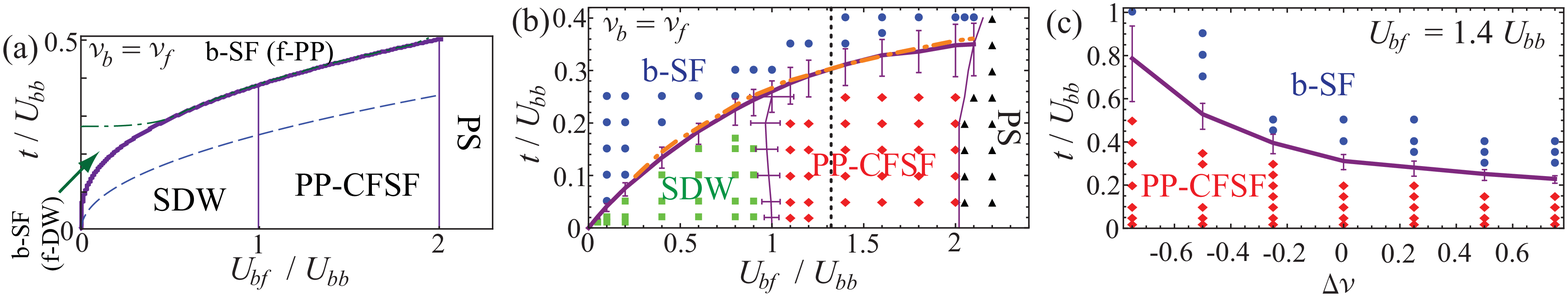}
\caption{\label{fig:PD}
(color online) Ground-state phase diagrams for the Bose-Fermi Hubbard model of Eq.~(\ref{eq:BFHH}) with equal hoppings, i.e., $t\equiv t_{b} = t_{f}$. The emerging phases include bosonic superfluid (b-SF), spin-density wave (SDW), polaron-paired counterflow superfluid (PP-CFSF), and phase separation (PS).
The phase diagrams from TLL theory (a) and TEBD (b) are depicted in the $(U_{bf}/U_{bb},t/U_{bb})$-plane, where $\nu_{b}=\nu_{f} = 0.5$. The purple thick-solid line represents the phase boundary between the two-component TLL and the Mott insulator, and the strong-coupling regime is achieved below the dashed line. The purple thin-solid lines separate the phases in the Mott insulating region. The green dot-dashed line in (a) separates the fermionic QLROs, i.e. polaron pairing (PP) and density wave (DW), in the two-component TLL region. The black dotted line in (b) represents $U_{bf}/U_{bb}=1.32$, corresponding to the $^{174}$Yb-$^{173}$Yb mixture of Ref.~\onlinecite{sugawa-11}. Our phase boundary is consistent with the results of Ref.~\cite{pollet-06}, shown as an orange dashed-dotted line.
(c) The phase diagram from TEBD is depicted in the $(\Delta \nu,t/U_{bb})$-plane, with $U_{bf}/U_{bb} =1.4$.
We take $L=80$ for the TEBD calculations.
}
\end{figure*}

In this Letter, we show that the PP-CFSF phase is naturally realized in Yb mixtures, which constitutes the first numerical demonstration of this phase in the BF-Hubbard model. In Fig.~\ref{fig:PD} we show the quantum phase diagram of the one-dimensional (1D) BF-Hubbard model, obtained with the quasi-exact time-evolving block decimation (TEBD) method~\cite{vidal-04}, supported by TLL theory. The Hamiltonian of the system is~\cite{albus-03}:
\begin{eqnarray}
H= -\sum_a\sum_{j} t_{a}(c_{a,j}^{\dagger} c_{a,j+1} + {\rm H.c.})
+ \sum_a\sum_{j} \epsilon_{j} n_{a,j} \nonumber \\
+ \frac{U_{bb}}{2}\sum_{j}n_{b, j}(n_{b, j}-1)
+ U_{bf} \sum_{j} n_{b,j} n_{f,j}, 
\label{eq:BFHH}
\end{eqnarray}
where $t_a$ is the tunneling energy of particle type $a = b, f$. $\epsilon_{j}$ denotes the external potential, and $n_{a,j}\equiv c_{a,j}^{\dagger} c_{a,j}$. $U_{bf}$ and $U_{bb}$ denote the on-site interactions. For a homogeneous system, we define the filling fractions $\nu_a \equiv \langle n_{a,j}\rangle$. Inspired by $^{174}$Yb-$^{173}$Yb mixtures reported in Ref.~\onlinecite{sugawa-11}, we assume that $U_{aa'} > 0$, $t \equiv t_{b}= t_{f}$, and $\nu \equiv \nu_f + \nu_b =1$. In the phase diagrams in Fig.~\ref{fig:PD},  we vary $t/U_{bb}$, $U_{bf}/U_{bb}$, and $\nu_{b} - \nu_{f}$. We find that the PP-CFSF phase occupies a broad regime between the SDW and PS regions, which is approximately $1<U_{bb}/U_{bf}<2$ and $t/U_{bb}<0.3$ when $\nu_{b}=\nu_{f}=0.5$. Since $U_{bf}/U_{bb}\simeq 1.32$ in $^{174}{\rm Yb}$-$^{173}{\rm Yb}$ mixtures, the ground state of the system is expected to be the PP-CFSF state. In the following, we first map out the ground-state phase diagram of the homogeneous system ($\epsilon_{j}=0$) varying 
  $\bar{t}\equiv t/U_{bb}$, $u\equiv U_{bf}/U_{bb}$, and $\Delta \nu \equiv \nu_b - \nu_f$ to find the PP-CFSF phase.  To connect closely to experiments, we also confirm that the PP-CFSF state emerges inside the Mott plateau with $\nu=1$ in the presence of a parabolic trapping potential.


We first determine the phase diagram via TLL theory. We go to a continuum representation, $c_{a,j}/d^{1/2}\rightarrow \psi_{a}(x)$,  where $d$ is the lattice constant, and express the  particle operators through a bosonization identity~\cite{haldane-81, cazalilla-04}:
\bea
\psi_{f/b} & = & [\rho_{f/b} + \Pi_{f/b}]^{1/2} \sum_{m\, odd / even} e^{i m \Theta_{f/b}} e^{i\Phi_{f/b}}, 
\eea
where $\rho_{f/b} = \nu_{f/b}/d$ are the average densities of the fermions/bosons, $\Pi_{f/b}(x)$ are the low-$k$ parts (i.e. $k\ll 1/\rho_f, 1/\rho_b$) of the density fluctuations, $\Phi_{f/b}$ are the phase fields, and $\Theta_{f/b}(x) = \pi \rho_{f/b} x + \theta_{f/b}(x)$, with $\theta_{f/b}(x)=\pi \int^x dy \Pi_{f/b}(y)$.
The right- and left-movers of the fermions are represented by $\psi_{f,R/L} \sim \rho_{f}^{1/2} e^{\pm i\Theta_{f}} e^{i\Phi_{f}}$. The action describing the mixture is 
$S  =  S_{0, f} + S_{0,b} + S_{bf} +S_{uk}$.
The terms $S_{0,a}$ are 
\bea
S_{0, a} & = & \frac{1}{2\pi K_{a}} \int d^2r \Big( \frac{1}{v_{a}}(\partial_{\tau} \theta_{a})^2 +  v_{a} (\partial_x \theta_{a})^2 \Big)
\eea
where $v_a$ and $K_a$ are the velocities and Luttinger parameters of the fermionic and bosonic liquid, for $U_{bf} =0$. We use $\hbar=1$ throughout the paper.
 The fermionic parameters are $v_f = 2 t d \sin (\pi \nu_f)$ and $K_f = 1$, the bosonic ones are $v_b = v_f (1 - 8 \bar{t} \nu_b \cos (\pi \nu_b))$ and $K_b = 1 + 8\bar{t} \sin(\pi \nu_b)/\pi$ assuming $\bar{t} \ll 1$. The interaction between bosons and fermions generates both linear terms, $S_{bf}$, and a non-linear term $S_{uk}$, in the effective action. $S_{bf}$ is 
\bea
S_{bf} & = &  \frac{U_{bf}}{\pi^2} \int d^2r \partial_x \theta_1 \partial_x \theta_2 
 + \frac{V_{bf}}{\pi^2} \int d^2r \partial_\tau \theta_1 \partial_\tau \theta_2.
\eea
The second term is created during the renormalization group (RG) flow; its prefactor therefore has the initial value $V_{bf}(0)=0$. The non-linear Umklapp term $S_{uk}$ is 
\bea
S_{uk} & = &  \frac{2 g_{uk}}{(2 \pi \alpha)^2} \int d^2r  
\cos(2 \theta_1 + 2\theta_2).
\label{Suk}
\eea
The renormalization group (RG) flow equations for this system were derived in Ref.~\cite{mathey-07-1}; its qualitative behavior is that of a Berezinskii-Kosterlitz-Thouless (BKT) transition~\cite{remark-1}. $S_{uk}$ can either be irrelevant, $g_{uk}\rightarrow 0$, or relevant, $g_{uk} \rightarrow \infty$. In Fig.~\ref{fig:PD}(a), the boundary between these regimes is depicted by the purple thick-solid curve. The relevant regime has two subregimes, namely the strong-coupling regime, in which $g_{uk}$ diverges rapidly, and the cross-over regime, in which $g_{uk}$ initially decreases. We estimate the boundary between these by the initial sign of the prefactor of the $g_{uk}$ flow equation (see Ref.~\cite{mathey-07-1}). Using the above estimates we find $U_{bf, c} = 16 t^2/U_{bb}(\sin(\pi \nu_b))^2 (1 - 4 (t \nu_b/U_{bb})\cos(\pi \nu_b))$, depicted by the blue dashed line in Fig.~\ref{fig:PD}(a). The orders indicated in this  figure are the dominant quasi-long-range order (QLRO), i.e. the order parameter $O(x)$ whose correlation function exhibits the slowest algebraic decay as $\langle O^{\dagger}(x) O(0)  \rangle \sim |x|^{-\alpha}$, with $\alpha < 2$, see e.g. Ref.~\cite{giamarchi-04}. 
In the regime where $S_{uk}$ is irrelevant, we find that  the most dominant QLRO is always the superfluid of bosons dressed with fermions whose order parameter is $O^{\rm DB}=\exp(-i\eta \Phi_{f})\psi_b$ with a real number $\eta$~\cite{mathey-04, mathey-07-2}. For the fermionic sector we find a competition of the $2 k_F$ component of the density operator of the fermions, $O^{\rm fDW}=\rho_f$, describing a spontaneous density modulation, and a polaron pairing operator $O^{\rm fPP}=\psi_{f, L} \psi_{f,R} \exp(i \lambda \Phi_b)$, with $\lambda$ a continuously varying parameter.  

For relevant $S_{uk}$, the total density is frozen out, which suggests a strong-coupling expansion leading to the following spinless Fermi-Hubbard model~\cite{kuklov-03, lewenstein-04},
\begin{eqnarray}
\!\!\! \hat{H}_{\rm f} \! = \! - J \! \sum_{ j} \! (f_j^{\dagger}f_{j+1} \! + \! {\rm H.c.})
\! + \! V \sum_{j} \! m_j m_{j+1}
\label{eq:hcb}
\end{eqnarray}
where $f_j \sim c_{f,j}c_{b,j}^{\dagger}$ is the polaron annihilation operator, consisting of an original fermion and a bosonic hole; $m_j \equiv f^{\dagger}_j f_j$, 
$V = 2(t_{b}^2 + t_{f}^2)/U_{bf} - (4t_{b}^2)/U_{bb}$, and 
$J = (2t_{b} t_{f})/U_{bf}$.
$f_j$ experiences no external field. The model of Eq.~(\ref{eq:hcb}) is solvable by Bethe ansatz, see e.g. Ref.~\cite{giamarchi-04}: For $V/J>2$ and $\nu_{p}\equiv \langle m_j\rangle=0.5$, i.e., ($0<$)$u<\frac{1}{2}(1-g)^2$ and $\nu_{b}=\nu_{f}$, where $g\equiv t_{f}/t_{b}$, N\'eel order with wave number $k=\pi/d$ emerges. This phase  is not  present when $t_{b}= t_{f}$. For $V/J <-2$, i.e. $u>\frac{1}{2}(1+g)^2$, strong attraction leads to the collapse of the polaron gas, i.e. phase separation of the original mixture. The remaining region is a TLL of  polarons.
For positive $V$, i.e., $u<\frac{1}{2}(1+g^2)$, the dominant QLRO is density wave ordering, corresponding to a SDW phase of the original mixture. For negative $V$, i.e. $u>\frac{1}{2}(1+g^2)$, the dominant QLRO is triplet polaron-pairing, i.e. the PP-CFSF phase, as indicated in Fig.~\ref{fig:PD}(a).
 
Having established this phase analytically, we now corroborate its existence numerically, and demonstrate its stability in a trapped system. 
We calculate the ground-state phase diagram using the TEBD method for open boundaries~\cite{vidal-04} via imaginary-time propagation. 
We fix the number of lattice sites to $L=80$. 

The TEBD method is a variant of the density-matrix renormalization group (DMRG), which allows to accurately calculate ground states of much larger systems than those tractable with exact diagonalization used in Ref.~\cite{roth-04}. In addition, it has the advantage that any correlation function can be calculated efficiently, in contrast to quantum Monte Carlo methods used in Refs.~\cite{takeuchi-05, pollet-06, zujev-08}, for example. Although DMRG has been applied to the BF-Hubbard model in Refs.~\cite{rizzi-08, mering-08}, the PP-CFSF phase has not been identified and explored, as we do here.

\begin{figure}[tb]
\includegraphics[scale=0.32]{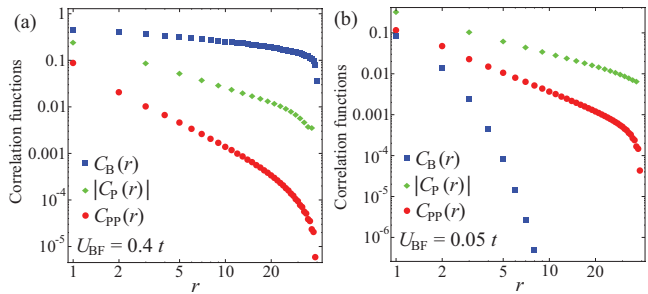}
\caption{\label{fig:corrs}
(color online) Blue squares, green diamonds, and red circles represent the correlation functions $C_{\rm B}(r)$, $|C_{\rm P}(r)|$, and $C_{\rm PP}(r)$, respectively,  for $L=80$, $U_{bf}/U_{bb} = 1.4$, and $t/U_{bb} = 0.4$ (a) and $0.05$ (b). The plots are on a log-log scale.
}
\end{figure}
\begin{figure}[tb]
\includegraphics[scale=0.33]{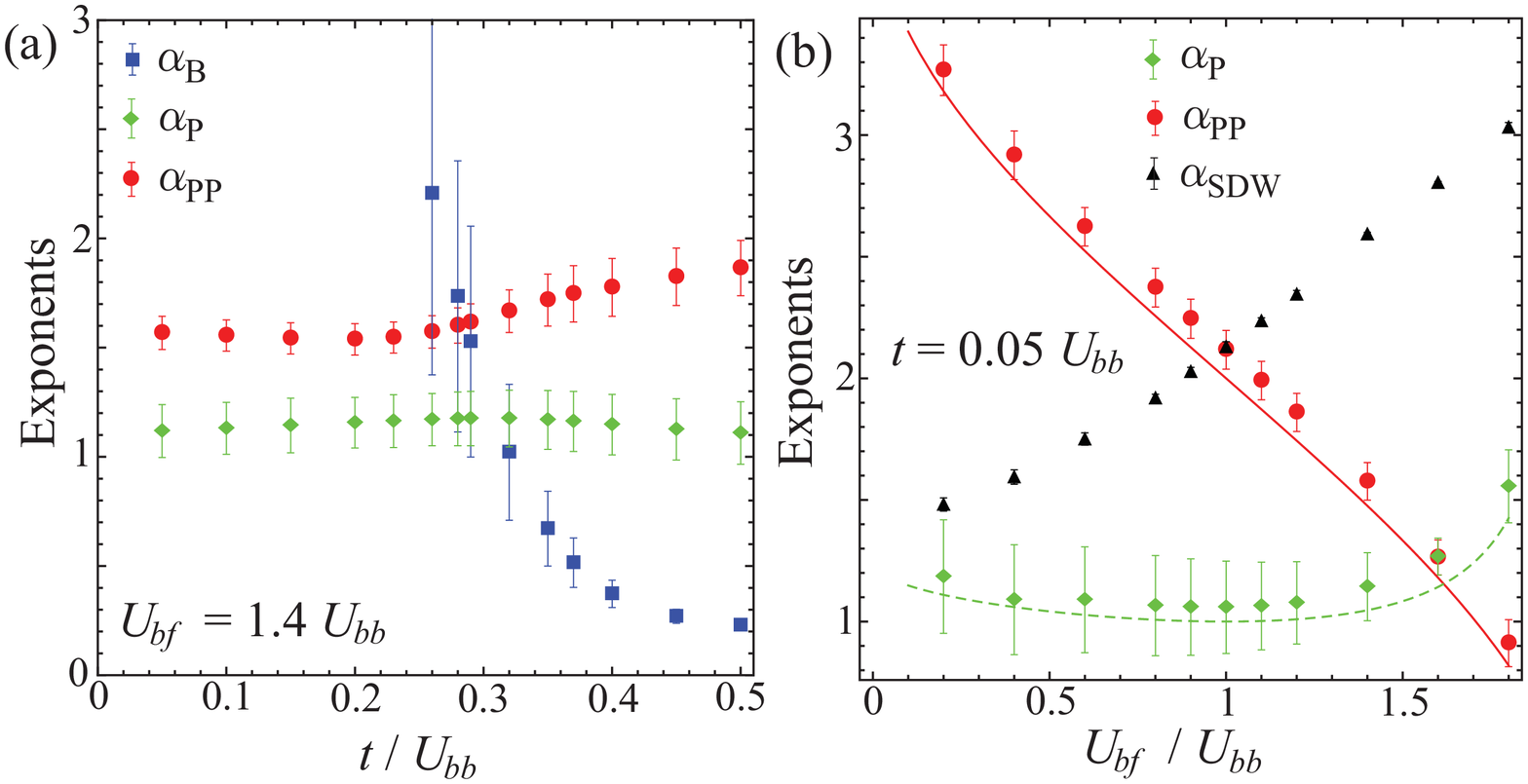}
\caption{\label{fig:expCorr}
(color online) The exponents  $\alpha_{\rm B}$ (blue squares), $\alpha_{\rm P}$ (green diamonds), $\alpha_{\rm PP}$ (red circles), and $\alpha_{\rm SDW}$ (black triangles). While $t/U_{bb}$ is varied at $U_{bf}/U_{bb}=1.4$ in (a), $U_{bf}/U_{bb}$ is varied at $t/U_{bb}=0.05$ in (b). The red solid and green dashed lines represent $\alpha_{\rm PP}=2/K_p$ and $\alpha_{\rm P}= K_p/2 +1/(2K_p)$, which are derived from TLL theory for the 1D polaron gas of Eq.~(\ref{eq:hcb})~\cite{giamarchi-04}. The Luttinger parameter is given by $K_p = \pi/(2\pi - 2 \cos^{-1}(1-u))$ at $\nu_p = 0.5$~\cite{cazalilla-04}. 
}
\end{figure}

We first determine the transition from the two-component TLL to the Mott insulator, at which the correlation function for the bosonic superfluid $C_{\rm B}(r)=\langle c_{b, h + r}^{\dagger} c_{b,h} \rangle$ switches from algebraic  to exponential decay. We choose $h \equiv L/2$. 
In Fig.~\ref{fig:corrs} we plot $C_{\rm B}(r)$, the polaronic correlation function $C_{\rm P}(r )\equiv\langle c_{b,h+r} c_{f,h+r}^{\dagger} c_{f,h}c_{b,h}^{\dagger} \rangle$ and the polaron-pair correlation function $C_{\rm PP}(r ) \equiv \langle (O_{r+h}^{\rm PP})^{\dagger} O_{r}^{\rm PP} \rangle$, where $O_{j}^{\rm PP}\equiv c_{f,j+1} c_{b,j+1}^{\dagger} c_{f,j} c_{b,j}^{\dagger}$.  $C_{\rm P}(r )$ decays algebraically and oscillates with the Fermi wave-number $k_f= \pi \nu_f/d$ as $C_{\rm P}(r) \sim \sin(k_f rd)|r|^{-\alpha_{\rm P}}$ at long distances.   $C_{\rm PP}(r) = \langle (O_{r+h}^{\rm PP})^{\dagger} O_{r}^{\rm PP} \rangle$ decays algebraically as $C_{\rm PP}(r) \sim |r|^{-\alpha_{\rm PP}}$. As shown in Fig.~\ref{fig:corrs}(a) (Fig.~\ref{fig:corrs}(b)), $C_{\rm B}(r)$ ($C_{\rm P}(r)$) decays more slowly in the two-component TLL (Mott insulating) regime. By fitting the correlation functions with $f(r)=\gamma|r|^{-\alpha}$, where $\alpha$ and $\gamma$ are fitting parameters, we extract the exponents $\alpha$ and plot them as functions of $\bar{t}$ in Fig.~\ref{fig:expCorr}(a). For small $\bar{t}$, the exponent for the bosonic correlation $\alpha_{\rm B}$ exceeds that of the polaronic correlation $\alpha_{\rm P}$ at what we define as the transition point. By this, we locate the Mott transition as shown by the purple thick-solid lines in Figs.~\ref{fig:PD}(b) and (c). The numerical phase boundary, for $L=80$, is closer to the strong-coupling regime determined by the RG analyses than the actual phase boundary, which is expected to emerge for larger systems. Note that the phase boundary agrees well with the results in Ref.~\cite{pollet-06}, depicted by the orange dashed-dotted line in Fig.~\ref{fig:PD}(b).

\begin{figure}[tb]
\includegraphics[scale=0.4]{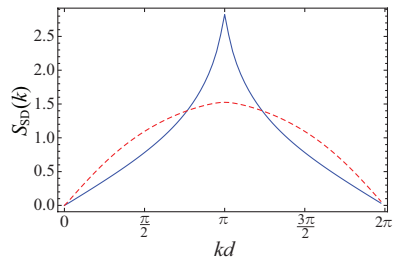}
\caption{\label{fig:Sq}
(color online) The static structure factor for the spin density $S_{\rm SD}(k)$ for $U_{bf}/U_{bb} = 0.4$ (blue solid) and $1.4$ (red dashed). We set $\nu_{b} = \nu_{f} = 0.5$ and $t/U_{bb} = 0.05$ at which the system is deep in the Mott insulating state.} 
\end{figure}

As discussed above, there are three phases in the Mott insulating regime, namely PS, PP-CFSF, and SDW, see~\cite{remark-2}. PS is signaled by an emerging peak in the bosonic structure factor 
 at low wave-number~\cite{pollet-06}. The PP-CFSF phase is characterized by the correlation function $C_{\rm PP}(r )$.  In Fig.~\ref{fig:expCorr}(a), we see that $\alpha_{\rm PP}<2$ in the Mott insulating regime at $u=1.4$, meaning that the system has PP-CFSF QLRO. To identify SDW QLRO, we calculate the static structure factor for the spin density 
\begin{eqnarray}
S_{\rm SD}(k) = \frac{1}{L}\sum_{j,l} 
(\langle \Delta n_j \Delta n_l \rangle - \langle \Delta n_j \rangle \langle \Delta n_l \rangle)
e^{-ikd(j-l)},
\end{eqnarray}
where $\Delta n_{j} \equiv n_{b,j} - n_{f,j}$. Since $\langle \Delta n_{h+r} \Delta n_{h} \rangle  \sim \cos(2k_f r d)|r|^{-\alpha_{\rm SDW}}$ at long distance, the structure factor behaves as $S_{\rm SD}(k) \sim ||k|-2k_f|^{\alpha_{\rm SDW}-1}$  near $|k|=2k_f$. Consequently, $S_{\rm SD}(k)$ has cusps at $k=\pm 2k_f$ when the SDW QLRO is present, i.e. $\alpha_{\rm SDW}<2$. In Fig.~\ref{fig:Sq}, we show $S_{\rm SD}(k)$ at $\nu_{b}=\nu_{f}=0.5$ and indeed find a cusp for $u=0.4$. By numerically fitting $S_{\rm SD}(k)$ near $|k|=2k_f$, we extract the exponent $\alpha_{\rm SDW}$. In Fig.~\ref{fig:expCorr}(b), $\alpha_{\rm SDW}$ is plotted as a function of $u$ for $\bar{t} = 0.05$, compared to $\alpha_{\rm PP}$. Clearly the system transitions from SDW to PP-CFSF when $u$ increases. 
As seen in Figs.~\ref{fig:PD}(b) and (c), the PP-CFSF phase 
is present approximately for $1 < u < 2$ and a wide range of $\Delta\nu$, consistent with the analytical results above. Note that in previous studies the polaron-pair correlation function was not considered, and thus the PP-CFSF phase not found.

\begin{figure}[tb]
\includegraphics[scale=0.29]{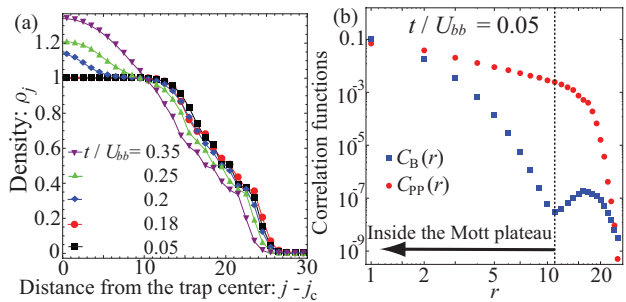}
\caption{\label{fig:DP}
(color online) Ground state properties in the presence of a parabolic trapping potential, where $N_b=N_f=20$, $\Omega/t = 0.008$, and $U_{bf}/U_{bb} = 1.4$.
(a) Density profiles $\rho_j$ for several values of $t/U_{bb}$ are shown. (b) The correlation functions $C_{\rm B}(r)$ and $C_{\rm PP}(r)$ are plotted on a log-log scale, where $t/U_{bb}=0.05$. 
} 
\end{figure}

Given that $u=1.32$ in the $^{174}$Yb-$^{173}$Yb mixture of Ref.~\onlinecite{sugawa-11}, PP-CFSF is expected to be realized in the dual Mott insulator regime when the  temperature is sufficiently low.
To demonstrate the stability of the phase in a trap, we calculate the ground states including a parabolic trap,
%
$\epsilon_{j} = \Omega (j-j_c)^2$,
%
where $\Omega$ and $j_c$ are the curvature and center of the trap. We set $\Omega/t =0.008$ and $N_b = N_f = 20$, where $N_a$ is the number of particle $a$. In Fig.~\ref{fig:DP}(a), we show the density profiles $\rho_j \equiv \langle n_{b,j}  +  n_{f,j} \rangle$ for several $\bar{t}$ while fixing $u=1.4$. At $\bar{t} = 0.35$, the system is compressible and no Mott plateau is present. When $\bar{t}$ decreases, a plateau at $\nu = 1$ is formed around $\bar{t} \simeq 0.18$. We calculate $C_{\rm B}(r)$ and $C_{\rm PP}(r)$ at $\bar{t}=0.05$ as shown in Fig.~\ref{fig:DP}(b). Inside the plateau, $C_{\rm B}(r)$ decays exponentially while $C_{\rm PP}(r)$ decays algebraically with the exponent $\alpha_{\rm PP} = 1.60 \pm 0.05 < 2$. Thus, there is PP-CFSF QLRO.

Finally, we estimate the temperature regime required for PP-CFSF to emerge. At finite temperatures ($T>0$) in 1D, the polaron-pair correlation function $C_{\rm PP}(r)$  exhibits  algebraic decay at $r \ll \xi$ while decaying exponentially for $r \gg \xi$, where the crossover length is $\xi = v_{p} / (\pi k_{\rm B} T)$ and the sound velocity of the polarons is $v_{p} = 2Jd\sin(\pi \nu_{p})(1 + V/(\pi J))$ for small $V/J$~\cite{cazalilla-04, giamarchi-04}. Hence, when $\xi \gg d$, PP-CFSF QLRO is considered to be sufficiently developed so that signatures of the PP-CFSF state can be observed. Since the condition $\xi \gg d$ corresponds to $J \gg k_{\rm B} T$, we estimate $J$, as an example, for $t/U_{bb} = 0.18$,  which is a modest value for a complete Mott plateau to be formed as shown in Fig.~\ref{fig:DP}(a). In the experiments of $^{174}$Yb-$^{173}$Yb mixtures, the lattice spacing is $d = 266 {\rm nm}$ and the s-wave scattering lengths are $a_{bb} = 5.55 {\rm nm}$ and $a_{bf} = 7.34 {\rm nm}$. Moreover, we assume the lattice depth in the transverse direction to be $V_{\perp} = 50 E_{\rm R}$, where $E_{\rm R}$ is the recoil energy. Using these parameters and the Wannier function obtained by numerically solving the Schr\"odinger equation with a sinusoidal potential, we estimate that $t/U_{bb} = 0.18$ is reached when the lattice depth in the axial direction is  $V_0 \simeq 2.33 E_{\rm R}$. At this lattice depth, $J \simeq k_{\rm B} \times 7.1 {\rm nK}$. Since the lowest temperature realized in Ref.~\cite{sugawa-11} is $\sim 5{\rm nK}$, the condition $k_{\rm B} T <  J$ is already possible with current experimental techniques. Further experimental advancement is necessary to achieve $k_{\rm B} T \ll J$.

In summary, we have used the time-evolving block decimation method and Tomonaga-Luttinger liquid theory to reveal the quantum phases of polarons inside the dual Mott insulator of one-dimensional Bose-Fermi mixtures at unit total filling. Interestingly, we found a large phase diagram regime with a counterflow superfluid phase of polaron pairs (PP-CFSF), in contrast to previous studies. We have shown that this state is expected to be formed in the dual Mott insulator regime of $^{174}$Yb-$^{173}$Yb mixtures in optical lattices at sufficiently low temperatures. 
\begin{acknowledgments}
The authors thank S. Sugawa, S. Taie, Y. Takahashi, and R. Yamazaki for enlightening discussions on their experiments with the $^{174}$Yb-$^{173}$Yb mixture. The computation in this work was partially done using the RIKEN Cluster of Clusters facility. LM acknowledges support from the Landesexzellenzinitiative Hamburg, which is financed by the Science and Research Foundation Hamburg and supported by the Joachim Herz Stiftung.
\end{acknowledgments}



\begin{thebibliography}{99}
\bibitem{remark-1}
This is in contrast to the universality of the analogous phase transitions in higher dimensions, discussed in Ref. ~\cite{altman-12}.
\bibitem{remark-2}
The QLROs in the two-component TLL are not studied here because our main focus is the PP-CFSF phase.
\end{thebibliography}
\end{document}